# Sham: A DSL for Fast DSLs

Rajan Walia[a], Chung-chieh Shan[a], and Sam Tobin-Hochstadt[a]

a   Indiana University, Bloomington, United States

**Abstract**   Domain-specific languages (DSLs) are touted as both easy to embed in programs and easy to optimize. Yet these goals are often in tension. Embedded or internal DSLs fit naturally with a host language, while inheriting the host's performance characteristics. External DSLs can use external optimizers and languages but sit apart from the host.

We present Sham, a toolkit designed to enable internal DSLs with high performance. Sham provides a domain-specific language (embedded in Racket) for implementing other high-performance DSLs, with transparent compilation to assembly code at runtime. Sham is well suited as both a compilation target for other embedded DSLs and for transparently replacing DSL support code with faster versions. Sham provides seamless inter-operation with its host language without requiring any additional effort from its users. Sham also provides a framework for defining language syntax which implements Sham's own language interface as well.

We validate Sham's design on a series of case studies, ranging from Krishnamurthi's classic automata DSL to a sound synthesis DSL and a probabilistic programming language. All of these are existing DSLs where we replaced the backend using Sham, resulting in major performance gains. We present an example-driven description of how Sham can smoothly enhance an existing DSL into a high-performance one.

When compared to existing approaches for implementing high-performance DSLs, Sham's design aims for both simplicity and programmer control. This makes it easier to port our techniques to other languages and frameworks, or borrow Sham's innovations "à la carte" without adopting the whole approach. Sham builds a sophisticated and powerful DSL construction toolkit atop fundamental language features including higher-order functions, data structures, and a foreign-function interface (FFI), all readily available in other languages. Furthermore, Sham's approach allows DSL developers to simply write functions, either using Sham or generating Sham, without needing to work through complex staging or partial evaluation systems.

## The Art, Science, and Engineering of Programming

 **Introduction**

Domain-specific languages (DSLs) have dual benefits.

1. By simplifying semantics and notation, they can make it easier for programmers to specify what they want without paying "attention to the irrelevant" [20].
2. By restricting freedom of expression they enable compilers to employ powerful optimizations otherwise ruled out by the flexibility of a general purpose language.

We see these dual benefits play out in everything from regular expressions, with dedicated syntax and specialized JIT compilers, to probabilistic programming languages, with mathematical semantics and automatic inference algorithms.

Unfortunately, it can also be a lot of work to build a DSL that brings about these benefits.

1. On one hand, implementing a DSL that simplifies semantics and notation requires taking care of concerns such as parsing, type checking, and runtime support. To this end, it is popular to take advantage of features of the *host language*—the general-purpose language the DSL is written in—such as Racket's macros, Haskell's type classes, or JavaScript's objects. Reusing host features can also serve to integrate the DSL with the host language and make it easier for them to call each other. Such interoperability can be achieved either by embedding the DSL as a library in the host language, or by implementing the DSL as a code generator (such as a Racket macro or a Template Haskell splice) that generates code in the host language. Indeed, high-level languages such as Racket are designed to host interoperating DSLs, by providing common reusable facilities such as a module system.
2. On the other hand, incorporating powerful optimizations into a DSL requires delving into details such as program analysis, code transformation, and memory management. To this end, reusing host features is easy but puts the DSL at the mercy of the host's performance. Instead, to reap the substantial benefits of domain-specific optimizations, the DSL implementation has to generate code in a low-level, potentially unsafe *target language*, such as C or LLVM. Building such a low-level code generator is tedious and error-prone, because the low-level language lacks high-level host facilities. It is also hard to integrate the generated code with high-level languages, in particular the host language. These difficulties are sadly well-represented in modern DSL design. For example, AugurV2 [10] generates C code that is then compiled and run, OptiML [4] generates C++ code that is then compiled and run, and Accelerate [18] developed a whole new binding to LLVM [17] to support integration with Haskell.

To offer an easier way to build high-performance DSLs there are frameworks that provide functionality for both high-level interface and low-level performance. These systems like Terra and LMS built using already existing popular languages modify the host to provide features for high performance compilers in an easy to use package. They provide a full blown package which requires big upfront commitment from DSL developers when deciding implementation techniques for their compilers.

This paper offers an easier way to build high-performance DSL compilers, by generating low-level code while still reusing and integrating with a high-level host language.

We design a compiler framework which provides convenience for building high-performance DSLs by providing the same benefits that make DSLs attractive:

1. Simple high level notation for language specification and code generation.

2. Optimization opportunities otherwise impossible in a high level host language.

We implement this framework without modifying the compiler or relying on multi stage type system but using DSL techniques itself. We implement a framework incremental in both use case and implementation complexity, our layered techniques don't require any modifications to the host language and provide independent "à la carte" ideas for other DSL framework authors.

Our framework **Sham** provides common facilities to support fast yet integrated DSLs. It builds upon the same language-oriented spirit in which existing high-level languages such as Racket provide support for embedded DSLs. In fact, Sham itself provides multiple embedded DSLs with domains in implementing high performance DSLs. Sham provides convenience for building embedded compilers similar to how a DSL improves solving a problem in its domain when compared to a general purpose language.

The design of Sham identifies four key components in a typical structure of a DSL compiler (especially embedded DSLs), pictured in Figure 1. Instantiating this general pattern, Figure 1b shows the typical design of a DSL in Racket: Racket macros enforce static semantics and generate expressions, which at runtime call specialized functions written to support the DSL [8]. The entire system relies on the Racket runtime, which provides services such as garbage collection, input-output handling, and integration with the wider operating system context. Although this design is powerful and convenient it bounds the resulting performance of a DSL by Racket's implementation. To remove this performance ceiling DSL implementations instead generate low-level code themselves to express domain-specific optimizations, as shown in Figure 1c. Unfortunately, the low-level code generator can no longer reuse host-language features such as variable binding, and the one-off integration built for a particular DSL compiler to coexist with the code it generates is usually ad-hoc and limited.

Figure 1d illustrates our new approach. Starting with the Racket approach in Figure 1b, we replace the generated code and runtime functions with Sham implementations—written in a lower-level language that generates low-level code transparently using LLVM. Significantly, our approach replaces DSL-generated programs with a second level of meta-programming: the DSL now generates Sham programs, which generate LLVM IR. However, the surrounding levels remain the same—code is still generated with Racket macros, and the entire system still relies on the Racket runtime.

This approach has three key benefits. First, it is incremental: one can replace even a single function with Sham code without interfering with the rest of the compiler. Second, it integrates smoothly: by keeping both the front-end and the underlying runtime, integration can be both closer and simpler than with a stand-alone program compiled to a different language. Third, it scales to a wide variety of DSLs: some need only a collection of functions, some need only code generation, and some need

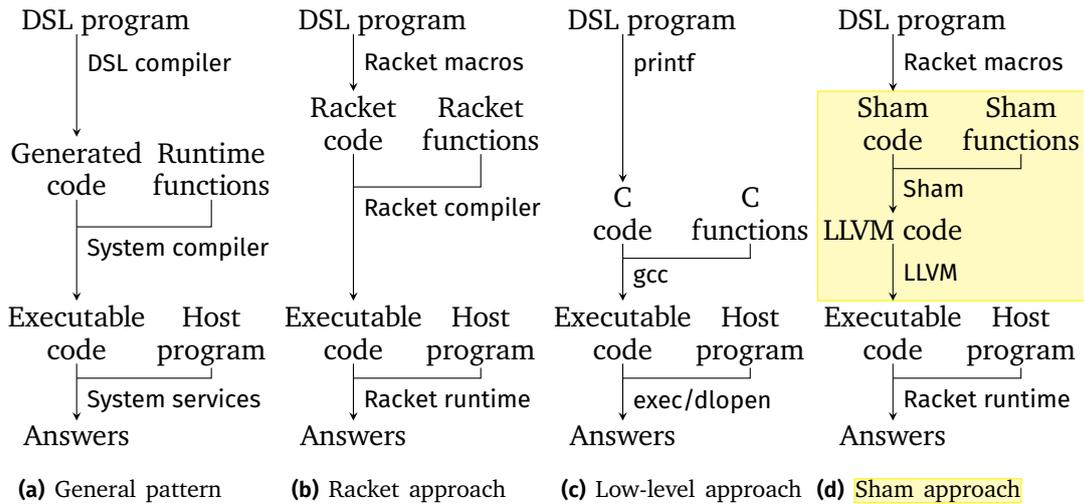

■ **Figure 1** Various ways to build a DSL

a comprehensive approach. We present examples of all three as case studies in this paper.

We designed Sham to serve two key use cases: writing it directly, as in a runtime function, and generating it programmatically, as in a macro-written front-end. We aim to make both uses of Sham straightforward, since both are crucial for the high-performance DSL author.

Additionally, to aid the DSL author, Sham comes with a library for automatically defining the abstract syntax and transformations needed in DSLs. This library `define-ast` proves its worth in our most complex case studies, as well as in Sham itself.

To introduce Sham, we begin in Section 2 with a simple example: finite state automata. This lends itself naturally to a simple macro that generates Sham code, producing a 2x speedup over the natural Racket implementation.

After describing the details of Sham in Sections 3 and 4, our next case study in Section 5 is a simple sound synthesizer. Here, the original DSL implementation uses macros to generate functions which call a collection of runtime functions, implemented in Racket. Porting these to Sham enables us to show the close correspondence between the original Racket code and the new Sham code. The performance improvement is up to an order of magnitude.

Last in Section 6, our most substantial case study is a new backend for the Hakaru probabilistic programming system. Using Sham, we replaced an existing Haskell-generating backend and implemented new optimizations with substantial performance benefits. The new system is up to 20× faster than before.

While Sham is a Racket-embedded DSL and pays close attention to the details of LLVM, our approach of using DSL techniques to provide—a dedicated low-level language target, tight and automatic integration, and a flexible representation of low-level programs, is more general. We anticipate that similar systems would work well in other DSL-building contexts with distinct backend technology, ranging from Scala to JavaScript and from CUDA to WASM.

## 2   Automata Language

We introduce Sham by building a small language for defining finite state automata. This example [12] shows the simplicity and benefits of using Sham to implement high-performance languages in Racket. We define a language for this automaton and compare its implementation using pure Racket and Sham. The similarity between the two shows that Sham is easy to adopt and achieves much faster performance.

We use this language to define the three-state automata from [12] recognizing the language c(a|d)*r, which includes words like car, cdr, cadr , and cddr, and so on:

```
(define-fsa M init (end)
  [init ([c more])] [more ([a more] [d more] [r end])] [end ()])
```

The code above serves as the input to both the Racket and Sham implementations. The starting state is init, and the final accepting states are (end). The state end has no outgoing transitions.

Our automata language is a pattern-matching macro in Racket:

```
(define-fsa name start (final ...) [state ([input next] ...)] ...)
```

This macro define-fsa takes the identifier name for the automaton, the identifier start for the initial state, the list (final ...) of final accepting states, and then a clause for each state. Each clause defines a state and its transitions. A transition is a pair of input and next. Racket does not differentiate between () and [] but they improve readability in places, here we use them to differentiate different groups in our macro syntax.

The result of compiling an automaton takes as an input a vector of symbols storing the word to match, and produces true or false as an output to indicate success or failure. Like [12], we implement this language by defining mutually recursive functions: for each state, we define a function, and for each transition, the old state makes a tail call to the new state. These functions take the input vector, the current position in the vector, and the total length of the vector, and do case analysis for each transition and call the function for that transition.

Compilation via Racket in Figure 2a iterates over the states and builds mutually recursive functions. We use syntax-parse [5], which can match the automaton program directly against the language syntax pattern above. The #:with clause binds the final result res of each state to true or false depending on whether state is in the final list or not. Inside we define a function for each state containing a case clause for each transition. We also generate a top-level function name for calling start state with initial inputs.

To compile through Sham, the first step is to determine the types. Because we are using symbols as the transition input, we use the type provided by Sham which mirrors Racket's symbols in a lower level. Sham's basic types closely reflect the type system of LLVM but it also provides types for Racket values. For the vector, we choose the type rkt-sym*, which is a pointer to an array of rkt-sym. Sham also provides utility functions to easily convert between Racket values and Sham values so we can convert a Racket vector of symbols to Sham's array using rkt-array-from-vector. The last input (len) is the length of the array. As we have no way to figure out the total length from a pointer, we include this extra argument in both the Racket and Sham

```
(define-simple-macro (define-fsa name start (final ...) [state ([input next] ...)] ...)
 #:with (res ...) (map (λ (e) (if (memq (syntax-e e) (syntax->datum #'(final ...)))
                                   #'true #'false))
                       (syntax->list #'(state ...)))
```

```
(begin                                  (define-sham-function
  (define                                 (name (inp : rkt-sym*) (len : i64) : rkt-bool)
    (name inp len)                         (return^ (start inp (ui64 0) len)))
    (start inp 0 len))                   (define-sham-function
  (define                                 (state (inp : rkt-sym*) (pos len : i64) : rkt-bool)
    (state inp pos len)                    (if^ (icmp-ult^ pos len)
    (if (< pos len)                           (switch^ (array-ref^ inp pos)
      (case (vector-ref inp pos)                [(rkt-sym input)
        [(input)                                 (return^ (next inp (add1^ pos) len))]
         (next inp (add1 pos) len)]               ...
         ...                                     (return^ (rkt-bool false))
         [else false])                       (return^ (rkt-bool res))) ...))
      res)) ...))
```

**(a)** Compiling through Racket      **(b)** Compiling through Sham

■ **Figure 2**   Comparing two different automata compilers implementations. Once we know the types, we add them to the function signature in a simple syntax Sham provides for defining a function.

The complete compiler through Sham, shown in Figure 2b, only requires changes as expected for a lower-level language. Racket's `if` changes to Sham's `if^`, and `case` changes to `switch^`. We change the `unsafe-vector-ref` to `array-ref^`, which does pointer arithmetic and then a load. We also have to provide explicit types for immediate integers like (`ui64 0`). When compiling an automaton, the identifier `input` in Figure 2b is a Racket symbol. But `inp` in the generated code is a Racket vector that stores symbols in Sham's representation, so we write a function `rkt-sym` to convert a Racket symbol to a Sham value.

Calling this macro on an automaton builds a collection of mutually recursive Sham functions defined using `define-sham-function`. This constructor only builds the syntax for Sham functions; they are not compiled yet. Compiling these functions in Sham requires an explicit step providing control over the compilation to the DSL writers. Sham compiles a set of functions together in a module. Invoking `define-sham-function` defines a function and attaches it to the module specified in the `current-sham-module` parameter. Once we have defined all the functions, we can compile the module by calling `compile-sham-module!`. After compilation, our `sham-app` operator can call a function from the current module. This looks up the compiled function and feeds it the given inputs after converting them using Racket's FFI library.

## 3   Sham

In this section we give a detailed overview of the Sham language interface and provide details about its implementation in Racket. Sham is a language designed for implementing high-performance DSLs in Racket. It is statically-typed similar to C and JIT-compiled using the LLVM toolchain, providing greater performance than what pure Racket code can achieve.

Sham is also an embedded language in Racket built using Racket's macro system and FFI library. Sham provides functions and macros to construct its syntax structures and compiles at run-time. Functions compiled using Sham are normal Racket functions. Sham ASTs are first-class Racket objects and constructed and combined using Racket expressions. Not only can the programmer call literal Sham constructors directly, they can also use Racket values to build specialized Sham code.

We use the `init` function from our automaton example in Section 2 to explain the Sham language. The Racket expression below builds Sham syntax for a function bound to the identifier `init`.

```
(define-sham-function
 (init (inp:rkt-sym*) (pos:i64) (len:i64) : rkt-bool)
 (if^ (icmp-ult^ pos len)
  (switch^ (array-ref^ inp pos)
   [(rkt-sym c) (return^(more inp (add1^ pos) len))]
   (return^(rkt-bool false)))
  (return^(rkt-bool false))))
```

### 3.1 Expressions

A Sham expression, similar to C expression, is a part of Sham function or statement and compiles to a piece of code that produces a value. Sham expressions are first-class objects in Racket, constructed using Racket functions and macros provided by Sham. The first-class property allows separate construction in different locations and later combined together. We begin with a small expression from our `init` example.

```
(more inp (add1^ pos) len)
```

This builds a Sham application whose operator is the Sham function `more` and arguments are inp, (add1^ pos), and len. Defining the Sham function `more` using `define-sham-function` also binds it as a Racket function that creates a Sham application expression when applied to Sham expressions. The Racket function call (add1^ pos) produces (add pos (ui64 1)), where add is a Sham operator for building integer addition and ui64 creates integer literals.

An abbreviated grammar for Sham expressions is:

| expr | ::= | *identifier* | \|(*identifier expr*…) |
|------|-----|--------------|------------------------|
| | \| | (app^ *rator expr* …) | \|(gep^ *expr expr*…) |
| | \| | (load^ *expr*) | \|(ui64 *integer type*) |
| | \| | (si64 *integer type*) | \|(let^ ((*identifier expr type*)…) *stmt expr*) |
| | | … | |

The simplest expression is a variable, which may be a function argument, a let^ variable, or a mutable variable. A let^ expression can bind multiple Sham variables in parallel; each variable has a name, an initial value, and also a type. Racket's syntax system tracks Sham variables and makes them available as Racket identifiers bound to Sham syntax. Sham expressions compile down to LLVM instructions and thus some directly correspond to specific LLVM IR, as with `gep^` ("get element pointer") which performs pointer arithmetic, and `load^` which reads from an address.

Sham has four different types of operators for building application expressions.

*rator* ::= *identifier* |(intrinsic *symbol type*)
| (external *symbol symbol type*) |(racket *symbol racket-value type*)

An identifier bound by `define-sham-function` can directly build a function application. LLVM intrinsics like `log`, `sin`, and `trunc` require their type specification. We can also use a function from a shared library by giving the `external` constructor the name of the library, the name of the function, and its type signature. Similarly, the `racket` constructor allows Sham's compiled code to call any Racket function. This powerful feature lets us switch back to Racket for parts of our DSL that are not yet compiled using Sham or that don't need Sham. This makes it easier to incrementally port functions to Sham while keeping the rest of the implementation unchanged, whereas compiling a DSL using an external language would require extra boilerplate code at each application.

Because each Sham syntax constructor is a normal Racket function, we can mix Racket and Sham seamlessly. Thus, we can build expressions for a complex DSL on the fly based on Racket values. One such example is to build a power function for a specific exponent *n*.

```
(define (build-pow x n)
  (if (= n 0) (ui64 1) (mul x (build-pow x (- n 1)))))
```
Here x is a Sham expression, n is a Racket number, and `mul` builds a multiplication in Sham. Thus, we use Racket to build a Sham AST with many multiplications. For example, the Racket call (`build-pow x 3`) builds the Sham expression (`mul x (mul x (mul x (ui64 1)))`) for cubing x.

### 3.2 Statements

Like C, Sham syntax not only has expressions but also statements. A statement represents an action executed for its side effect. Like expressions, statements in Sham are also first-class Racket values. To explain Sham statements, we look at one from our running example of the `init` function.

```
(if^ (icmp-ult pos len)
    (switch^ (array-ref^ inp pos)
      [(rkt-sym c) (return^ (more inp (add1^ pos) len))]
      (return^ (rkt-bool false)))
    (return^ (rkt-bool false)))
```
Whereas `if` is Racket syntax, `if^` is a syntax for building Sham statement from a Sham expression (the condition) and two Sham statements (the "then" and "else" branches). Thus, as we have seen, we sometimes use caret `^` to distinguish Sham syntax from Racket.

Along with Racket syntax for generating Sham AST we also provide constructor functions which take syntax values as input arguments. For example, `switch^` is a macro whose syntax is analogous to Racket `case`, whereas the constructor for switch takes an expression (scrutinee), a list of expression-statement pairs (cases), and a statement (default). The macro is helpful when writing switch statement directly and the function is helpful for programmatically building case statements.

$$
\begin{array}{rll}
stmt & ::= & expr & |(\texttt{block}^\wedge \ stmt \ \ldots) \\
     & | & (\texttt{svoid}^\wedge) & |(\texttt{return}^\wedge \ expr) \\
     & | & (\texttt{while}^\wedge \ expr \ stmt) & |(\texttt{if}^\wedge \ expr \ stmt \ stmt) \\
     & | & (\texttt{set!}^\wedge \ var \ expr) & |(\texttt{switch}^\wedge \ expr \ (expr \ stmt) \ \ldots \ expr) \\
     & & \ldots
\end{array}
$$

Sham has the usual statements expected in a lower-level language. An expression used as a statement for side effect can be an application of a function performing series of actions. `while^` and `if^` follow those in C. The `set!^` statement is an assignment and mutates local variables introduced by functions and `let` bindings.

### 3.3 Functions

Sham provides two ways (`define-sham-function` and `sham-function`) of creating functions with Racket macros. Both macros create Sham syntax for a function, but `define-sham-function` in addition binds the function name to the created syntax and registers the function in the current global module. Racket identifier bound using `define-sham-function` performs the task of containing Sham syntax, constructing Sham application nodes and later looking up the compiled form. It is both a Racket value containing the Sham syntax, as well as a Racket procedure that creates a Sham application expression when applied to other Sham syntax values. Our previous automata example in Figure 2b shows the function more (more generally next) directly building a function application when applied to Sham expressions.

```
(define f (sham-function (identifier (identifier : type) ... : type) stmt))
(define-sham-function (identifier (identifier : type) ... : type) stmt)
```

The `sham-function` macro provides a way for creating functions at local context. This constructs a function syntax object locally and needs attaching to a module for compilation. We can also unquote the name of the function in `sham-function` to dynamically specify function names. For example, the code in Listing 1 uses `sham-function` to manually generate Sham functions named `more-0`, `more-1`, ..., to match the language $(ad)^{\texttt{len}-1}r$ for a specified `len`. As this code illustrates, the macro `sham-function` requires the name of a function, the names and types of the arguments, the return type, and a function body statement. Optional attributes attached using the `#:md` keyword are extra information required for specific optimizations. For example, `always-inline` forces LLVM to inline the function.

### 3.4 Modules and Compilation

A module is the smallest unit of compilation in Sham, consisting of function definitions, type definitions and mutable variables. All the functions compiled together in the

**Sham: A DSL for Fast DSLs**

■ **Listing 1** Generating a series of Sham functions

```
(define fmd (sham-function-metadata #:attributes '[always-inline]))
(define (build-mores len)
  (define names (for/list ([i len]) (format-symbol "more-~a" i)))
  (cons
   (sham-function #:md fmd
    (,(list-ref names (sub1 len)) (inp : rkt-sym*) : rkt-bool)
    (switch^ (array-ref^ inp (ui64 len))
      [(rkt-sym r) (return^ (rkt-bool true))]
      (return^ (rkt-bool false))))
   (for/list ([i (sub1 len)] [name names] [next (cdr names)])
     (sham-function #:md fmd (,name (inp : rkt-sym*) : rkt-bool)
       (switch^ (array-ref^ inp (ui64 len))
         [(rkt-sym a) (return^ (app^ next inp))]
         [(rkt-sym d) (return^ (app^ next inp))]
         (return^ (rkt-bool false)))))))))
```

■ **Listing 2** A complete example of using Sham

```
(define-current-sham-env pow-module)
(define-sham-function (pow (x : i64) (n : i64) : i64)
 (if^ (icmp-ule^ n (ui64 0)) (return^ (ui64 1))
  (return^ (mul^ x (pow x (sub-nuw^ n (ui64 1)))))))
(sham-jit-compile! pow-module #:opt-level 3)
(sham-app pow 2 10)
```

same module can refer to each other by name, as well as to any type defined in the module, providing support for defining mutually-recursive functions and providing more optimization opportunities.

Listing 2 shows a full example of using Sham, demonstrating not only the language for writing low-level code, but also operations to control the compilation of defined code. Here `define-current-sham-env` is syntax for specifying an environment for Sham compilation: when defining a function using `define-sham-function` it is by default added to the module stored in the current environment. Specifying a module explicitly gives more control and is useful in case of multiple modules defined together.

The function `sham-jit-compile!` compiles all the functions defined in a module. It takes an environment and an optional optimization level using the `#:opt-level` keyword argument. The optimization level is similar to `-O1`, `-O2`, `-O3` in C compilers. The programmer can also exert finer control over the compilation and optimization by running a specific pass from the LLVM's pass library on the module.

After compiling a module, we can look up the compiled version of a function either by name or using the identifier with `sham-lookup-function`. Sham also provides default wrapper for these raw compiled functions as Racket functions: Racket's FFI library allows us to wrap a C-level function into a Racket function by providing the

```
<definition>   ::=   <id> <group>+              top level definition
<group>        ::=   <id> <production>*          a set of productions
<production>   ::=   <id> <pattern>
<pattern>      ::=   `single' <id>              group or production label
                |   `repeat' <pattern>
                |   `multiple' <pattern>*
                |   `terminal' <id>            racket value type
```

■ **Figure 3** Grammar for defining ast structures in Racket

input types and return type. Once compiled `sham-app` can apply these functions to Racket values of corresponding types. Moreover, it is possible to get the raw pointer address of a compiled function, which is useful when we are defining a different module and we want to call the previously compiled function without incurring the overhead of calling through Racket.

### 3.5 Language Syntax

Sham comes with a small DSL for defining Racket's structure types for storing syntax trees useful for general DSL implementations. This small language specifies a language AST with the grammar in Section 3.5 which then expands to Racket code generating the boilerplate structures and functions.

```
(define-ast LC
 (expr [lambda ((x:expr.sym ...) body:expr)]
       [letrec (((ids:expr.sym vals:expr) ...) body:expr)]
       [app (rator:expr rand:expr ...)]
       [n #:terminal number?]
       [sym #:terminal symbol?]))
```

The macro `define-ast` specifies syntax information for a language. It requires a name for the language and a context free grammar specification in the form of groups and nodes. A group is a set of alternative productions each building a Racket structure type. Node productions store syntax following their pattern which can reference other productions, a repetition, or a sequence of values. The `single` pattern follow the format `<variable-name>:<production-reference>`, where `<variable-name>` is the field name for underlying structure. The above example produces the following structure for the `letrec` node; defining a structure `expr-letrec` as a child of `expr` containing three fields `ids` (a list of `sym`), `vals` (a list of `expr`), and `body` (an `expr`).

```
(struct expr-letrec expr [ids vals body])
```

The language specification can also build functions to traverse over the generated AST structures shown in Listing 6 or pretty printers. We make use of generic methods for Racket structures to implement an overloaded method for each node. Generic methods for structures in Racket lets us define a function with same name but different definition for each node. Pretty printers helps focus on only the important parts of syntax values and are useful for debugging while implementing a DSL compiler.

We define Sham's AST itself using `define-ast` and also provide it as a library for other language developers to use, showing its use in implementing our example language Hakaru in Section 6.

## 4 Sham's Implementation

Sham's goal is an easy to use library for implementing high performance just-in-time compilers for DSLs written in Racket. To achieve this we focus on the following implementation ideas.

- A library for an alternate compiler without any modifications to Racket's language implementation. We achieve this by making use of the FFI library and syntax utilities provided by Racket.
- Multiple compatible interfaces for ease of use in both programmatic code generation and handwritten code. We provide this by making use of `define-ast` and Racket's language building libraries.
- Separating LLVM library interface and a sufficiently higher level intermediate language for compilers. We achieve this by dividing our implementation in two major sections.
- Competent interface for interacting with Racket with minimum boilerplate code. We make use of Racket's FFI and embedding support for providing this interaction.

We expand on these ideas and their incorporation in the following subsections.

### 4.1 Pure Racket Library

One big reason for the popularity of a library is its ease of use. Requiring changes to a language runtime increases friction. Sham provides a complex alternate compiler library without requiring any changes to Racket. This is possible due to the features Racket provides by default to its users for building language extensions. This means installing Sham is only a single command away and we do not have to maintain a separate implementation of our host language Racket.

### 4.2 User Interface

Sham provides two interfaces for writing programs, one aimed for code generation and the other for human writing low-level code by hand. A DSL implementation consists of both a code generator and a language library of standard functions and data structures. In a high-performance DSL implementation that compiles using a low-level language, the library—unlike the code generator—is usually implemented in C and linked at compile time. This approach disconnects the library from the compiler and invites code duplication. To improve this situation, Sham provides not only a functional interface for code generation but also a syntactic interface. Using Sham, the compiler implementer can write all the code without leaving Racket, keeping the code base concise and also improving performance optimizations like inlining.

Sham's AST is itself built using the `define-ast` form and comes with all the features described in Section 3.5. This includes the availability of functional constructors for Racket structures, pattern matching on language constructs, and traversing and mapping operations.

Sham follows Racket's design of using macros to expand complex forms to a set of core constructs. For example, the core construct for constructing a Sham function is `sham:def:function`, which requires a name, type, and body statement. This form is simple and concise but tedious to use. To ease the tedium, Sham also provides `sham-function`, which expands to the correct function type constructor, binds functions arguments in the body, and stores the name of the function in language syntax. A higher-level form `define-sham-function` expands to `sham-function` and moreover registers the functions to the current global environment for compiling and defines a Racket value that can be directly used to create Sham application syntax. These syntax forms at different levels of detail provide the user with a choice at each use site. Also, Sham's syntax macros conveniently provide constructs that combine the definition, compilation, and invocation of a set of functions. The details are kept usually implicit using thread-local parameters but still available if required.

### 4.3 Distributing Complexity

The Sham language described in Section 3 compiles to LLVM in two steps. First, Sham is lowered to a language representing LLVM IR in Racket. Second, LLVM's C library is used. This separation of concerns not only distributes the complexity of implementing Sham but also provides the LLVM IR as another target language besides Sham.

**Sham's LLVM IR**   LLVM's original C-API is an imperative builder interface that requires a user to maintain state and build a module by calling provided C functions to add instructions one by one. This is a tedious and error-prone process duplicated across multiple implementations of DSLs targeting LLVM. To reduce this duplication, and also to simplify Sham's design, we designed a language that constructs LLVM IR in memory. This language serves as a target for Sham.

This language closely mirrors LLVM's IR and provides a functional interface to LLVM's C-API. Building LLVM IR in memory also removes disk IO when compiling at run-time. As with Sham's language syntax, this language is also implemented using `define-ast` and provides similar interface for generating language terms.

**Compiling to LLVM IR**   Sham contains language constructs that are not available as LLVM IR instructions and that compile to a series of LLVM instructions. Constructs requiring lowering to multiple instructions include nested expressions and statements, function-local variables and assignments, and referencing fields in a structure.

Each instruction in LLVM IR is in static single assignment (SSA) form and thus variables cannot change their values. We compile Sham variables not to LLVM variables but to LLVM stack references, and then optimize the loads and stores using LLVM's optimization pass for global value numbering. This approach allows us to not worry about phi nodes in SSA form and still have complex branch instructions. We similarly

■ **Listing 3** Automata compiled using LLVM basic blocks and jumps

```
(define-sham-function
  (name (inp : rkt-sym*) (pos : i64) (len : i64) : rkt-bool)
  (label^ state
    (if^ (icmp-ult^ pos len)
        (switch^ (array-ref^ inp pos)
          [(rkt-sym input) (set!^ pos (add^ pos (ui64 1)))
                                (label-jump^ next)]
          ...
          (return^ rkt-false))
        (return^ res))) ...)
```

handle the complexity of compiling function arguments and let-bound variables to LLVM IR. We also need to flatten expressions and statements in Sham for compiling to LLVM IR, and convert `while` and `if` statements to LLVM basic blocks and jumps.

**Directly constructing LLVM IR**   Sham implements its constructs on top of LLVM IR, but the use of basic LLVM operations is also sometimes needed. Instead of adding all the LLVM IR operations to Sham, we designed the two languages to work together by allowing higher-level Sham to embed the LLVM representation (similar to assembly blocks in C). This way, we keep the best of both worlds without overly complicating either of these languages. Listing 3 illustrates this embedding by using low-level basic blocks and jumps to implement our `cadr` automata from Figure 2. Here `label^` creates a basic block named `state`, and `label-jump^` jumps to the basic block named `next`. This definition would replace the uses of `define-sham-function` in Figure 2b.

The code generated by Figure 2b is simple enough for LLVM to optimize the tail calls to direct jumps automatically. The resulting code is similar, though not identical, to that generated by Listing 3, with no measurable run time performance impact. However, in more complex cases, where LLVM's optimizer may be less successful, directly constructing jumps offers the ability to ensure the desired low-level behavior.

### 4.4  Interacting with Racket

Sham provides an extended set of types and functions for interacting with Racket at run-time. Sham implements this interaction with the Racket virtual machine using Racket's FFI library and internal functions provided for embedding support. Coexistence of low-level code compiled through LLVM with Racket values and functions allows the language designer to implement only part of their compiler in Sham and get performance benefits from LLVM. Sham provides flexible support for three aspects of this coexistence; how best to use these options depends on DSL requirements, which vary by a great deal in different implementations.

**Communicating values with Racket**   Running Sham code inside Racket runtime alongside Racket functions requires transfer of data at boundaries of Sham and Racket.

There are two options for this transfer but their performance depends on each use-case. When accessing data fields once or twice it is better to avoid the upfront cost of complete conversion and to just use wrapper functions. However, it is beneficial to convert the whole structure once if Racket code will access values often. To cater both of these situations we provide support for both options in Sham.

- Convert the data structures from Racket's internal format to low-level C-style structures.
- Keep Racket's internal representation across both Sham and Racket functions.

Converting data representation requires copying bits and allocating extra memory but comes with other benefits. Keeping native representation means already existing code does not need any changes and when using external libraries it becomes necessary to use this approach. However, depending on the situation, if results are short lived it could be excessive to do this conversion. To reduce the complexity of memory allocation in these situations Sham can also use Racket's garbage collector to use managed memory.

Where the first option has an upfront cost at the boundary, the second option requires accessing data in a non-native representation and performance cost at each access. To support this approach we use Racket's internal functions like `scheme_make_pair` and `scheme_make_struct_instance` for constructing and manipulating Racket data structures so there is no extra overhead for using them in Sham. Racket also provides high level function to access C data structures as part of it's FFI library that can access Sham values. Presence of mutation complicates the situation and converting the whole structure is not feasible when modifying data directly.

When using Sham there is some performance cost for communicating values but our performance evaluation and case studies have shown that overall benefits outweigh these costs and with above mentioned two different approaches we can minimize the performance hit.

**Calling Sham from Racket**    Sham's lower-level LLVM IR compiles functions to raw pointer addresses, and the higher-level API provides functionality for wrapping these functions into Racket functions. By overridable default, these wrapped functions automatically convert values based on their types. Using some advanced features of Racket's FFI library, this conversion allocates output pointers automatically, calculates function arguments like array length, and attaches deallocators to returned values.

**Calling Racket from Sham**    There are two ways to call Racket functions from inside Sham, either by using an application expression with the `racket` operator introduced in Section 3.1, or by declaring a function at module level. Both of these options require a full type of the Racket function for converting values at the boundary. This makes it easier to start using Sham with parts of implementation still in Racket and also provides a simple way to debug lower-level code.

## 5 Synth

In this section we demonstrate how we can incrementally port a Racket DSL to use Sham and get a high-performance implementation. We use an already existing small DSL for sound synthesis and show the performance improvements obtained by using Sham.

*Synth* is a language for generating sound from a sequence of music notes and drum patterns. In *Synth* a sound is a list of audio samples generated using a specification of signals. *Synth* is a small language which deals with generating sound as a list of numbers (audio samples), while Racket is really good at providing an interface for creating such a language. Its virtual machine lacks options for optimizing programs generating large arrays of numbers. We show how we used Sham to only change parts needing performance improvements while keeping the rest of code unchanged and thus having a high-performance implementation along with Racket's high level language building ecosystem.

**Original implementation**  The original implementation of Synth from [1] parses the audio specification using Racket's syntax macros and expands it directly into Racket functions. It generates the audio sample and saves it into a file in `wav` format. The generated audio samples in the form of a list of numbers combine and mix together based on the specification. Each signal function generates a list of samples by using wave generators or combining sub signals.

**Improved implementation**  We improve upon the original implementation by staging our evaluation and preallocating the final vector and thus removing the overhead of allocation. In the original implementation every time we generate samples for a note and create a sequence we allocate a new array. Mixing signals combines the values of two arrays to generate a new array containing the result. This allocation of arrays for every operation costs a lot of time.

We achieve another boost in performance by splitting our synthesis in two steps. We emulate this in Racket by first calculating the number of samples, note frequencies and generating closures which takes a vector as an input. In the next step we run these closures on the preallocated vector. This staging brings us a step closer to using Sham as we only change the first step to use Sham.

**Further improving with Sham**  Sham provides an opportunity for further performance improvements with minimal changes to this Racket implementation. We show how we utilized the simple design of Sham to use it in an ad-hoc fashion by compiling only the performance critical parts and getting the most performance with minimal effort.

We compare the implementation of a function before and after modifications done for using Sham. The original implementation of sawtooth is in **??** 4a; we transform it into Sham function shown in **??** 4b. A quick glance shows that they are similar; we just have to provide more information in Sham version. It requires explicit types for function arguments, explicit type coercions and even immediate values like `1.0` need

■ **Listing 4**   Sawtooth wave generator through Racket and through Sham

```
(define (sawtooth-wave freq x)
 (let ([period (round (/ (sampling-freq) freq))]
       [period/2 (quotient period 2)]
       [x* (exact->inexact (modulo x period))])
   (- (/ x* period/2) 1.0)))
```

**(a)** Generator through Racket

```
(define-sham-function (sawtooth-wave [freq:f32] [x:i32]:f32)
 (let ([period (ri^ round.f32 f32
                    (fdiv (fl32 (sampling-freq)) freq))]
       [period/2 (ri^ trunc.f32 f32 (fdiv period (fl32 2.0)))]
       [x* (frem (ui->fp x (etype f32)) period)])
   (return (fsub (fdiv x* period/2) (fl32 1.0)))))
```

**(b)** Generator through Sham

to have an explicit type. Another difference is that Sham distinguishes expressions from statements and requires a return statement at the end.

The biggest difference between two is that one is a Racket *function* whereas the other is a Racket value representing a Sham function. Functions `fsub` and `fdiv` generate ASTs using the given arguments, just as `return` creates an AST for returning. We therefore separate generation, compilation, and execution of Sham code.

In the Sham version of the sawtooth wave generator, we can see three variables bound by Racket's `let`. They bind Racket-level variables to Sham ASTs, rather than creating an AST for Sham's `let^`. Another important thing to note is that (`sampling-freq`) is a Racket expression which is obtaining sampling frequency parameter. This is different than other expressions as we are calling a Racket function to calculate a number and then using that number in our code generation. It shows how we can combine Racket expressions with Sham expressions when constructing Sham AST to seamlessly perform partial evaluation.

**Advantages of Sham**   Sham is a low level programming language efficient at crunching raw numbers; while Racket is good at syntax macros and pattern matching on lists. In our improved implementation we separate the execution in two parts, the first doing pattern matching on lists to compute signal frequencies and note values and the second to calculate signal values. The first step runs once for a small amount of total time, whereas the second part has tight loops which run for significant amount of total time. This means even implementing the second part in Sham improves the overall performance drastically. Implementing only the performance-critical parts in a low level language provides performance improvements with less effort.

Sham also provides options to optimize the generated code—we use these tuning parameters to inline small functions and embed array addresses into the generated code. This makes use of information only available at the time of generating Sham code when compiling for a specific song.

We compare performance of two implementations of Synth: the improved Racket implementation, and the high-performance just-in-time compiled implementation using Sham. The Sham implementation is orders of magnitude faster than the original implementation. In our largest example Sham takes 38ms to compile and 109ms to run while (improved) Racket took 2888ms. Using Sham not only makes up for the overhead of run-time compilation but also improves the total running time by an order of magnitude. More details are available in Appendix C.

## 6  Hakaru

We used Sham as the backend code generator of a probabilistic programming language, and implemented a battery of low-level yet domain-specific optimizations succinctly by taking advantage of functions generated using `define-ast`. Sham helped us to achieve speed-ups by at least one order of magnitude, compared to an old backend that emitted Haskell code and compared to other hand-tuned, specialized Java implementations of similar languages.

The probabilistic language Hakaru [19, 26] expresses computationally intensive sampling algorithms at a high level, especially those that iterate over large arrays. Initial compiler passes use symbolic mathematics [25] to turn the high-level description of an algorithm into a side-effect-free program with nested loops over numerical arrays. The latter program, in the *Hakaru IR* language, gets compiled to machine code using Sham after domain specific optimizations implemented using `define-ast`.

We define the AST of Hakaru IR using `define-ast`, which represents side-effect-free programs with nested loops over numerical arrays. Listing 8 shows this definition written with Sham.

### 6.1  Domain-Specific Optimizations

We make use of functions generated by `define-ast` to transform the AST of Hakaru IR in optimization passes. These optimization passes are worth their while because the typical Hakaru program is short but computationally intensive: the end application may need to execute the same screenful of code for seconds or hours. Their implementation becomes more concise and maintainable by using AST constructors and generic map combinator, especially when Hakaru syntax evolves. Listing 9 shows an example optimization pass written using the Sham language definition framework.

Our most important transformations on Hakaru IR are conversion to A-normal form and loop-invariant code motion (LICM). LICM is especially profitable in this domain because the initial IR contains many loop-invariant subexpressions that are themselves loops; for example, normalizing an array means dividing each element by the sum of all the elements. Since Hakaru IR is pure, we know that our compiler preserves semantics as it performs LICM aggressively.

After transforming Hakaru IR, we lower Hakaru IR into Sham and perform additional optimizations there. During lowering, we fuse together loops that iterate over the same index bounds and take the same scope. Fusion during lowering is easy to implement because the source language Hakaru IR is side-effect-free whereas the target language Sham supports mutating the accumulators of multiple fused loops.

Another optimization we perform on Sham IR is to specialize the run-time generated code to input sizes and array addresses. This optimization is especially profitable because the sampling algorithms we target (Markov Chain Monte Carlo) spend most of their time repeatedly traversing large arrays of the same size. Thus we allow the user to provide binding-time information such as static size for array arguments or even static value, and we propagate such static data to intermediate arrays. Since Sham compiles to LLVM at run-time, the application can wait until input arrives to specialize the compilation with such static data, and doing so is profitable because the resulting machine code is computationally intensive. We give LLVM constant-size array types, such as `[10 x i64]` rather than `i64*`, to help it unroll loops. In the common case where constant propagation renders the size of an intermediate array static, we even allocate the intermediate array before LLVM compilation and inline its pointer address into the generated code, thus reducing the total register pressure. This optimizations is possible only due to Hakaru's semantics and its general use case and general purpose compilers like LLVM cannot perform these optimizations out of box. Sham plays an key part in this optimization as it would not be possible to perform this optimization without Sham's access to memory addresses of low-level data structures in a high-level language like Racket.

Sham helps implement all these optimizations and reduces boilerplate code. These opimizations must be implemented in our domain-specific setting, instead of relying on a general-purpose compiler, because the profitability and safety of the optimizations are hard to establish in a general-purpose language: loop complexity is hard to determine, and LLVM expressions may have side effects. In contrast, most loops iterate over large arrays of fixed size, and Hakaru IR has no side effects.

## 6.2 Performance

We evaluate our backend on the *collapsed Gibbs* sampling algorithm [14, 15] for the Dirichlet-multinomial Naive Bayes topic model [16, 22], comparing four implementations: our compiler backend based on Sham; our compiler backend without run-time specialization; a previous Hakaru backend that emits Haskell code; MALLET, a specialized system for statistical natural-language processing whose handwritten code we configured to run the same algorithm.

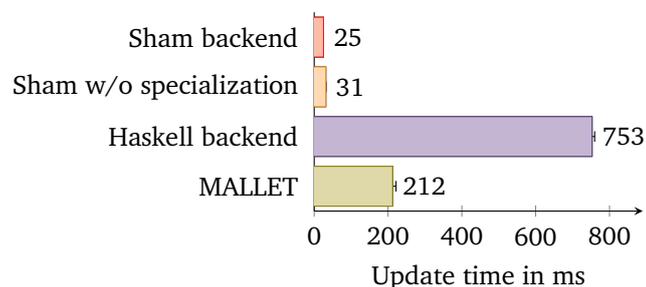

**Figure 4** Run time per Gibbs update in the topic model

Because these are implementations of the same algorithm, we omit comparison of accuracy and focus on running time. Further detail on the benchmarks is available in D.3 Figure 4 plots the time it takes for each system to compute a single Gibbs update. The results show that code generated by the Sham backend is an order of magnitude faster compared to code generated by the Haskell backend and compared to MALLET, and 24%+ faster than without specialization.

To measure the effect of our domain specific optimizations on the final performance we perform an ablation study. Section 6.2 shows the run time with different optimizations disabled. Although these optimizations have a combined effect, this table show how individual optimizations affect overall performance. We were able to implement these optimizations conveniently in a high level language Racket by making use of Sham's infrastructure.

■ **Table 1** Run time in seconds (mean over 1000 trials and standard error) of one sweep of Gibbs sampling with $m = 50$ and $n = 10000$. Slowdown with respect to full optimization.

| Optimizations | Time in seconds | Slowdown |
|---|---|---|
| No optimizations | 471.4 ± 0.6 | 1848 × |
| No histogram | 460.6 ± 0.2 | 1805 × |
| No LICM and loop fusion | 328.7 ± 0.1 | 1289 × |
| No loop fusion | 0.471 ± 0.003 | 1.8× |
| No run-time specialization | 2.422 ± 0.005 | 9.5× |
| Full optimization | 0.255 ± 0.001 | — |

## 7 Related Work

The performance benefits available to DSLs has spurred many language designers to develop approaches related to Sham. The most closely related systems are those that also generate low-level code for easy use from a high-level language, and we begin by relating Sham to two such systems, LMS and Terra summarized in Table 2. We then describe the relationship to other systems that share some goals and/or techniques with Sham but are not full systems with the same aims.

Terra [6, 7], a low-level systems programming language embedded and meta-programmed in Lua [9], is the most closely related system. While the two systems are technically similar, Sham differs from Terra most significantly in overall goal: Terra aims to allow programmers to write individual high-performance programs, whereas Sham aims to help DSL authors. Sham thus differs in several ways to facilitate easier development of DSL compilers and focuses less on direct use of the language.

1. *Language not API:* Sham provides an API to a compiler in a DSL format whereas Terra is an embedded language used directly.

2. *Opaque ASTs:* In both Terra and Sham syntax for functions, expressions and types are first class values in host language but in Sham syntax constructors are first class values exposed as normal Racket functions. These constructors let the compiler programmer to programmatically specify function names, function arguments etc which is not possible in Terra. Also, Sham users can analyze and deconstruct its AST values supporting expressive metaprogramming.

**Table 2** Comparison of systems for dynamic generation of low-level code

|                      | Sham                  | LMS                     | Terra                    |
| -------------------- | --------------------- | ----------------------- | ------------------------ |
| Target               | LLVM                  | Scala/C/OpenCL/Cuda     | C/LLVM                   |
| Primitives           | Functions and Macros  | Type based overloading  | Special syntax           |
| Value Conversion     | Explicit cast         | Optional cast           | Automatic conversion     |
| Complex data types   | C-style structure types | Scala data structures | C-style structure types  |
| Memory operations    | Pointers, malloc, free | Scala level operations | Pointers with Lua's GC   |
| Syntactic forms      | Racket macros         | Syntax Overloading      | Special syntax           |
| Stage polymorphism   | No (not wanted)       | Type-directed           | No                       |
| Transform ASTs       | Yes                   | No                      | No                       |
| Portability          | Specialized to LLVM   | Backend agnostic        | Tied to Clang            |
| Compilation          | Compiles at run-time  | Outputs code to file    | Compiles at run-time     |
| Modification to host | No modifications      | Language virtualization | New syntax, runtime      |

3. *Implicit staging:* Sham has a bright-line distinction between Sham values and Racket values; even integers must be explicitly converted to Sham integers, making it possible to combine Sham code with explicit quotation and unquotation. In Terra, values are automatically converted from Lua. Explicit conversion gives control to the compiler writer but requires more work when writing Sham directly.

4. *Limited control of compilation:* Terra allows both ahead-of-time and just-in-time compilation using a simple function call. Sham provides an extensive API to provide the compiler writer with greater control, allowing compilation of a group of possibly mutually recursive functions together using LLVM's *modules*. Sham also allows DSL developers to control compilation details of optimization pass selection, specifying LLVM attributes, etc.

LMS [23] is a library for multi-stage programming in Scala. It uses Scala's type classes to overload operators used with staged and unstaged code. While LMS is also aimed at accelerating DSLs, it is typically used to generate specialized code to be compiled and run later, not to be loaded dynamically by the generating programs.

1. *Type based overloading:* LMS uses Scala's typeclass functionality to distinguish between staged and unstaged code, whereas Sham uses Racket's syntactic forms to do the same. Sham makes a value conversion explicit for even basic types whereas LMS uses Scala's typeclasses to perform this conversion which is simpler for basic types but harder for complex types.

2. *Interpreters to compilers:* LMS aims at upgrading interpreters of domain specific languages written in Scala to use multi stage programming for performance whereas Sham aims at DSLs designed as straightforward compilers to Racket switching to use just-in-time compilation for performance.

3. *Target agnostic:* LMS is target agnostic and can compile to different backends whereas Sham is specific to LLVM. This has both pros and cons, making LMS a higher level language whereas Sham's focus on a single target allows it to provide compilation and performance options which are not possible in LMS. For example, Sham meta-programs can control naming choices whereas LMS automatically generates names, making debugging more difficult. Sham also aims at JIT compilation and execution, whereas LMS supports targets which are separately compiled and provides less support for JIT approaches.

4. *High level language:* LMS provides a similar level of abstraction to Scala itself, automatically transforming Scala code with type-embedded stage information into specialized and potentially lower-level code. In contrast, Sham gives the DSL developer more control, asking them to write lower-level code directly, and supporting functionality like manual memory allocation and pointer arithmetic which is important in developing high performance compilers.

5. *Modified compiler:* LMS, like Terra, uses a modified version of Scala's compiler to add virtualized operators whereas Sham requires no modification to Racket.

DSLs and DSL frameworks based on LMS have demonstrated the potential of this approach. Forge [24] is a framework for generating DSL implementations using a specification. Forge has its own high-level specification language, from which an implementation is automatically generated. This approach allows for building DSLs quickly by giving up low-level control. Delite [4] is a compiler framework and runtime for parallel embedded DSLs but does not include a language for building DSLs. Spiral [21] is a program generator for linear digital signal processing transforms.

AnyDSL [13] is a partial evaluation framework for programming high-performance libraries. Sham focuses on different goals when it comes to implementing high performance DSLs. Sham is for directly writing a compiler embedded in Racket whereas AnyDSL generates a compiler from partial evaluation.

1. *Partial evaluation:* AnyDSL's approach to DSLs is to build a compiler by partially evaluating an interpreter. Sham does not provide automatic partial evaluation and relies on the compiler implementor to perform any needed optimizations.

2. *No language embedding:* AnyDSL framework has its own language (Impala) which directly targets the Thorin IR and thus the DSLs produced are not embedded in a host language but are standalone compilers.

RPython [2] is a subset of Python programming language used to implement languages as simple interpreters. RPython generates interpreters and virtual machines for high level dynamic programming languages by automatically inserting appropriate low-level aspects. Sham on the other hand is a low level language used to write run-time compilers for DSLs embedded in Racket.

1. *Interpreters to compilers:* RPython generates a high performance tracing JIT compiler automatically from an interpreter, whereas in Sham the language implementer has to write the compiler.

2. *No language embedding:* The virtual machine produced by RPython is a standalone language implementation and is not embedded in any other host language, whereas Sham's most important goal is to embedding in a host language.

3. *No low-level control:* RPython is a subset of Python that allows little low-level control over memory allocation and pointer manipulation since it re-uses Python's garbage collector.

Cython [3] is a super-set of Python extending it with support for calling C functions and declaring C types for variables. The compiler is able to generate efficient C code based on this information. Cython aims for a language for wrapping external C libraries and embedding in other applications. As implementing DSLs is not a direct goal it lacks in utilities Sham provides specifically for implementing other languages. Cython also performs ahead of time compilation whereas Sham focuses on JIT compilation.

*LLVM Bindings:* There are other examples of language libraries exposing LLVM API differently. For example, the Haskell LLVM bindings, developed as part of Accelerate, provide a typed interface to LLVM [17]. Sham is a language abstraction which is simpler and easier than the API while providing control over LLVM-level details such as optimization, along with the language Sham provides tooling for building runtime compilers and interacting with host language. It not only provides the LLVM library as a language but also provides streamlined approach to using LLVM as a JIT compiler.

## 8    Conclusion

We present Sham, a low-level DSL embedded in Racket. The target domain of Sham is *implementing other high performance DSLs*. We designed Sham based on the insight that DSLs embedded in Racket require generating code and calling custom functions, both of which are performance sensitive and thus call for a high level approach to performance-oriented DSL construction.

Sham's approach to compiler construction makes it easy to *programmatically generate* low-level code as well as *directly author* programs with low-level control. This approach shows support for high performance DSL implementations does not require complicated type directed staging or creating a separate programming language. We can provide similar levels of convenience as other DSL frameworks with common language features by designing our programming interface incrementally.

We validate Sham with case studies: an implementation of Krishnamurthi's pedagogical automata DSL, a simple existing sound synthesis DSL in Racket, and Hakaru, an existing probabilistic programming language with a Haskell backend. In all three, we took an existing DSL, wrote both Sham code and Sham code generators, and achieved significant performance improvement, ranging from 2× to 20×. The results show that Sham is an effective choice for performance-oriented embedded DSLs.

**Acknowledgements**    This research was supported by the National Science Foundation under Grant No.1823244 and Grant No.1763922.

## References

[1]    Vincent St-Amour. *Simple software synthesizer in Racket*. https://github.com/stamourv/synth. 2013.

[2]    Davide Ancona, Massimo Ancona, Antonio Cuni, and Nicholas D. Matsakis. "RPython: A Step towards Reconciling Dynamically and Statically Typed OO

Languages". In: *Proceedings of the 2007 Symposium on Dynamic Languages*. DLS '07. Montreal, Quebec, Canada: Association for Computing Machinery, 2007, pages 53–64. ISBN: 9781595938688. DOI: 10.1145/1297081.1297091.

[3]   Stefan Behnel, Robert Bradshaw, Craig Citro, Lisandro Dalcin, Dag Sverre Seljebotn, and Kurt Smith. "Cython: The Best of Both Worlds". In: *Computing in Science & Engineering* 13.02 (Mar. 2011), pages 31–39. ISSN: 1558-366X. DOI: 10.1109/MCSE.2010.118.

[4]   Kevin J. Brown, Arvind K. Sujeeth, Hyouk Joong Lee, Tiark Rompf, Hassan Chafi, Martin Odersky, and Kunle Olukotun. "A Heterogeneous Parallel Framework for Domain-Specific Languages". In: *Proceedings of the 2011 International Conference on Parallel Architectures and Compilation Techniques*. PACT '11. 2011. DOI: 10.1109/PACT.2011.15.

[5]   Ryan Culpepper and Matthias Felleisen. "Fortifying Macros". In: *Proceedings of the 15th ACM SIGPLAN International Conference on Functional Programming*. ICFP '10. Baltimore, Maryland, USA: Association for Computing Machinery, 2010, pages 235–246. ISBN: 9781605587943. DOI: 10.1145/1863543.1863577.

[6]   Zach DeVito and Pat Hanrahan. "The Design of Terra: Harnessing the Best Features of High-Level and Low-Level Languages". In: *1st Summit on Advances in Programming Languages (SNAPL 2015)*. Volume 32. Leibniz International Proceedings in Informatics (LIPIcs). 2015, pages 79–89. ISBN: 978-3-939897-80-4. DOI: 10.4230/LIPIcs.SNAPL.2015.79.

[7]   Zachary DeVito, James Hegarty, Alex Aiken, Pat Hanrahan, and Jan Vitek. "Terra: A Multi-Stage Language for High-Performance Computing". In: *Proceedings of the 34th ACM SIGPLAN Conference on Programming Language Design and Implementation*. PLDI '13. Seattle, Washington, USA: Association for Computing Machinery, 2013, pages 105–116. ISBN: 9781450320146. DOI: 10.1145/2491956.2462166.

[8]   Matthias Felleisen, Robert Bruce Findler, Matthew Flatt, Shriram Krishnamurthi, Eli Barzilay, Jay A. McCarthy, and Sam Tobin-Hochstadt. "A programmable programming language". In: *Commun. ACM* 61.3 (2018), pages 62–71. ISSN: 0001-0782. DOI: 10.1145/3127323.

[9]   Luiz Henrique de Figueiredo, Roberto Ierusalimschy, and Waldemar Celes. "Lua: An Extensible Embedded Language". In: *Dr. Dobb's Journal* 21.12 (1996), pages 26–33.

[10]  Daniel Huang, Jean-Baptiste Tristan, and Greg Morrisett. "Compiling Markov Chain Monte Carlo Algorithms for Probabilistic Modeling". In: *SIGPLAN Not.* 52.6 (June 2017), pages 111–125. ISSN: 0362-1340. DOI: 10.1145/3140587.3062375.

[11]  Thorsten Joachims. "A Probabilistic Analysis of the Rocchio Algorithm with TFIDF for Text Categorization". In: *ICML '97: Proceedings of the 14th International Conference on Machine Learning*. Edited by Douglas H. Fisher. Nashville: Morgan Kaufmann, 1997, pages 143–151. ISBN: 1-55860-486-3.

[12]   Shriram Krishnamurthi. "EDUCATIONAL PEARL: Automata via macros". In: *Journal of Functional Programming* (2005). DOI: 10.1017/S0956796805005733.

[13]   Roland Leißa, Klaas Boesche, Sebastian Hack, Arsène Pérard-Gayot, Richard Membarth, Philipp Slusallek, André Müller, and Bertil Schmidt. "AnyDSL: A Partial Evaluation Framework for Programming High-performance Libraries". In: *Proc. ACM Program. Lang.* OOPSLA (2018). DOI: 10.1145/3276489.

[14]   Jun S. Liu. "The Collapsed Gibbs Sampler in Bayesian Computations with Applications to a Gene Regulation Problem". In: *Journal of the American Statistical Association* 89.427 (1994), pages 958–966. DOI: 10.1080/01621459.1994.10476829.

[15]   Jun S. Liu, Wing Hung Wong, and Augustine Kong. "Covariance Structure of the Gibbs Sampler with Applications to the Comparisons of Estimators and Augmentation Schemes". In: *Biometrika* 81.1 (1994), pages 27–40.

[16]   Andrew McCallum and Kamal Nigam. "A Comparison of Event Models for Naive Bayes Text Classification". In: *AAAI-98 workshop on learning for text categorization*. Volume 752. 1998, pages 41–48.

[17]   Trevor L. McDonell, Manuel M. T. Chakravarty, Vinod Grover, and Ryan R. Newton. "Type-safe Runtime Code Generation: Accelerate to LLVM". In: *Proceedings of the 2015 ACM SIGPLAN Symposium on Haskell*. Haskell '15. 2015. DOI: 10.1145/2887747.2804313.

[18]   Trevor L. McDonell, Manuel M.T. Chakravarty, Gabriele Keller, and Ben Lippmeier. "Optimising Purely Functional GPU Programs". In: *Proceedings of the 18th ACM SIGPLAN International Conference on Functional Programming*. ICFP '13. 2013. DOI: 10.1145/2544174.2500595.

[19]   Praveen Narayanan, Jacques Carette, Wren Romano, Chung-chieh Shan, and Robert Zinkov. "Probabilistic Inference by Program Transformation in Hakaru (System Description)". In: *FLOPS*. Volume 9613. Lecture Notes in Computer Science. Springer, 2016, pages 62–79. DOI: 10.1007/978-3-319-29604-3_5.

[20]   Alan J. Perlis. "Epigrams on Programming". In: *SIGPLAN Notices* 17.9 (1982), pages 7–13. DOI: 10.1145/947955.1083808.

[21]   Markus Püschel, José M. F. Moura, Jeremy Johnson, David Padua, Manuela Veloso, Bryan W. Singer, Jianxin Xiong, Franz Franchetti, Aca Gačić, Yevgen Voronenko, Kang Chen, Robert W. Johnson, and Nick Rizzolo. "SPIRAL: Code Generation for DSP Transforms". In: *Proceedings of the IEEE* 93.2 (2005). Special issue on program generation, optimization, and platform adaptation, pages 232–275. DOI: 10.1109/JPROC.2004.840306.

[22]   Philip Resnik and Eric Hardisty. *Gibbs Sampling for the Uninitiated*. Technical report CS-TR-4956 UMIACS-TR-2010-04 LAMP-TR-153. University of Maryland, June 2010. URL: http://www.umiacs.umd.edu/~resnik/pubs/gibbs.pdf.

[23]   Tiark Rompf and Martin Odersky. "Lightweight Modular Staging: A Pragmatic Approach to Runtime Code Generation and Compiled DSLs". In: *Proceedings of the Ninth International Conference on Generative Programming and Component Engineering*. GPCE '10. 2010. DOI: 10.1145/1868294.1868314.

[24]   Arvind K. Sujeeth, Austin Gibbons, Kevin J. Brown, HyoukJoong Lee, Tiark
       Rompf, Martin Odersky, and Kunle Olukotun. "Forge: Generating a High Perfor-
       mance DSL Implementation from a Declarative Specification". In: *"Proceedings
       of the 12th International Conference on Generative Programming: Concepts &
       Experiences"*. GPCE '13. 2013. DOI: 10.1145/2637365.2517220.

[25]   Rajan Walia, Praveen Narayanan, Jacques Carette, Sam Tobin-Hochstadt, and
       Chung-chieh Shan. "From High-level Inference Algorithms to Efficient Code".
       In: *Proc. ACM Program. Lang.* 3.ICFP (July 2019), 98:1–98:30. ISSN: 2475-1421.
       DOI: 10.1145/3341702.

[26]   Robert Zinkov and Chung-chieh Shan. "Composing Inference Algorithms as
       Program Transformations". In: *UAI*. AUAI Press, 2017.

## A   Automata Details

We include a full version of `more` function generated by `define-fsa` macro in both
Racket and Sham implementations in Listing 5.

▪ **Listing 5**   Two different outputs of `define-fsa` for `more` function.

```
(define (more inp pos len)
  (if (< pos len)
      (case (unsafe-vector-ref inp pos)
        [(c) (more inp (add1 pos) len)]
        [(a) (more inp (add1 pos) len)]
        [(d) (end  inp (add1 pos) len)]
        [else false])
      false))
```

**(a)** Racket version

```
(define-sham-function
 (more (inp:rkt-sym*) (pos:i64) (len:i64) : rkt-bool)
 (if^ (icmp-ult^ pos len)
  (switch^ (array-ref^ inp pos)
   [(rkt-sym c) (return^(more inp (add1^ pos) len))]
   [(rkt-sym a) (return^(more inp (add1^ pos) len))]
   [(rkt-sym d) (return^(end inp (add1^ pos) len))]
   (return^(rkt-bool false)))
  (return^(rkt-bool false))))
```

**(b)** Sham version

■ **Listing 6**   Code generated by `define-ast` for `letrec` production

```
(define-generics exprg (map-expr f-expr exprg))
(struct expr-letrec expr (ids vals body)
 #:methods gen:exprg
 ((define (map-expr f-expr exprg)
    (match-define (expr-letrec ids vals body) exprg)
    (expr-letrec (map f-expr ids) (map f-expr vals)
      (f-expr body)))))
```

■ **Listing 7**   Constant folding using the generated `map-expr`

```
(define (constant-fold e)
  (match (map-expr constant-fold e)
    [(expr-app (expr-sym '+) `(,(expr-n v) ...))
     (expr-n (apply + v))]
    [e e]))
```

## B   Language Definition

Listing 7 shows an example of using the `map-expr` generic method generated through `define-ast` (shown in Listing 6) to implement a constant folding pass for `+` and `*` applications. This `constant-fold` function takes an expression and recursively folds sum and multiply operations over constant values. To achieve this we call the `map-expr` generated by Sham with the function (`constant-fold`) and the input expression. In the body we use Racket's match operation to pattern match on the result of `map-expr`. We match for application nodes where the operator is either `+` or `*` and all the operands are numbers. The pattern (`list (expr-n v) ...`) makes sure all of the operands are of form `expr-n` and binds `v` to the list of numbers. We then `apply` either sum or multiply to the list of numbers and then wrap them in `expr-n` syntax construct. The match body performs constant folding at one level and `map-expr` applies it on the whole tree.

## C   Synth Performance

We compare the two implementations of Synth language, the improved Racket implementation, and the high-performance just-in-time compiled implementation using Sham. We measure the runtime for three songs in Figure 6, omitting Racket startup time and disk I/O. Here the total time for the Sham implementation is the sum of compile time and run time columns. The Sham implementation is more than 5× faster in every case. Additionally, the cost of compilation is re-gained by the benefits of optimization. We also measure total run time with increasing input size in Figure 5 to measure the performance benefits for longer running calculations, our results show Sham performs even better for bigger input size.

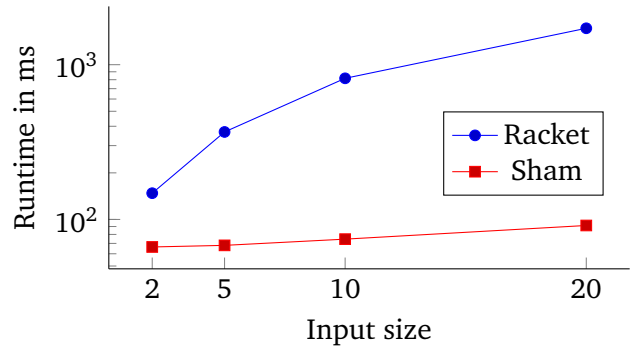

■ **Figure 5** Benchmarking total run-time of funky town with increasing input size.

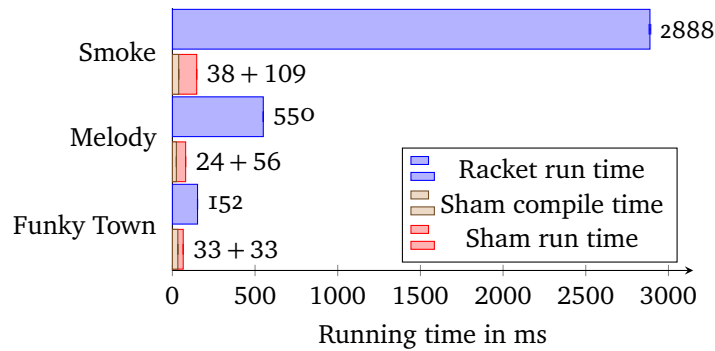

■ **Figure 6** Benchmarking running time of Synth using pure Racket implementation and new Sham implementation. Some standard errors are too small to show.

## D Hakaru

### D.1 AST with `define-ast`

We define the syntax of Hakaru IR by a grammar with three top level groups: `expr` (expression to produce a value), `reducer` (reducer to produce a histogram), and `pat` (pattern to match against and bind variables). Listing 8 shows a selection of productions for each of these groups. The original 50 line definition expands to more than 200 lines of Racket code which we initially wrote by hand.

### D.2 Hakaru Optimizations

The built-in operators and looping constructs are the locus of our compiler optimizations. We express our optimizations using the structure definitions, pattern-matching macros, and combinators generated by `define-ast`. The forms `create-pass` and `create-rpass` define top-down and bottom-up rewriting rules by non-exhaustive pattern-matching. We construct these macros using the `map` operation generated by `define-ast`. Listing 9 shows three bottom-up rewriting rules for array access. The first rewrites an `indexing` expression from $a[$if $e_0$ then $e_1$ else $e_2]$ to

■ **Listing 8**  Hakaru IR AST defined using `define-ast` form from Sham (excerpt)

```
(define-ast hakaru
 (expr [val (type v)]
       [if (type tst:expr thn:expr els:expr)]
       [app (type rator:expr rands:expr ...)]
       [bucket (type s:expr e:expr r:reducer)]
       [match (type tst:expr branches:expr ...)]
       [branch (p:pat body:expr)]
       [intrf (sym)]
       [var (type sym info)])
 (reducer [(index (n:expr i:expr a:reducer))]
          [(nop ())])
 (pat [(pair (a:pat b:pat))]))
```

■ **Listing 9**  An optimization pass on Hakaru IR, defined by bottom-up rewriting rules (excerpt)

```
(create-rpass
 (expr
  [(expr-app ta (expr-intrf 'index)
    `(,(expr-app t (expr-intrf 'array-literal) contents)
      ,(expr-if t chk (expr-val 'nat vthn) (expr-val 'nat vels))))
   #:when (< (length contents) 5)
   (expr-if t chk (list-ref contents vthn)
                  (list-ref contents vels)]
  [(expr-app ta (expr-intrf 'index) `(,(expr-var tv vv info) ,ind))
   #:when (and (list? tv) (is-constant-type? (second tv)))
   (expr-val ta (get-constant-value (second tv)))]
  [(expr-app ta (expr-intrf 'index)
    `(,(expr-app t
                 (expr-intrf 'constant-value-array)
                 `(,size ,content))
                 ,ind))
   content])
 (reducer)
 (pat)))
```

if $e_0$ then $a[e_1]$ else $a[e_2]$ where $a$ is a literal array. The other two rules simplify indexing into an array with values known at compilation.

### D.3 Hakaru Performance

Measurements in Figure 4 exclude the time Sham takes to generate and compile the Gibbs update code: $416.9 \pm 0.4$ ms with specialization and $468.6 \pm 0.3$ ms without specialization. This cost is more than made up for by the speed of the generated code, because even just a single sweep of Gibbs sampling requires running the generated code for a single update 2000 times.

This benchmark uses the 20 Newsgroups corpus [11]. We hold out 10% of the classifications and infer them. All benchmarks ran on a 6-core AMD-Ryzen 5 with 16 GB of RAM and Linux 4.15. We used Racket 6.12, LLVM 5.0.1, and GHC 8.0.2.

Additional evaluation of the Hakaru system presented in [25].

## About the authors

**Rajan Walia** rawalia@indiana.edu

**Chung-chieh Shan** ccshan@indiana.edu

**Sam Tobin-Hochstadt** samth@indiana.edu